# Spectral statistics of rare-earth nuclei: Investigation of shell model configuration effect

H. Sabri*

Department of Physics, University of Tabriz, Tabriz 51664, Iran.

* Author E-mail: h-sabri@tabrizu.ac.ir

## Abstract

The spectral statistics of even-even rare-earth nuclei are investigated by using all the available empirical data for *Ba*, *Ce*, *Nd*, *Sm*, *Gd*, *Dy*, *Er*, *Yb* and *Hf* isotopes. The Berry- Robnik distribution and Maximum Likelihood estimation technique are used for analyses. An obvious deviation from GOE is observed for considered nuclei and there are some suggestions about the effect due to mass, deformation parameter and shell model configurations.

## Introduction

The microscopic many-body interaction of particles of Fermi systems such as heavy nuclei is rather complicated. Therefore, several theoretical approaches to the description of the Hamiltonian which are based on the statistical properties of its discrete levels are applied for solutions of realistic problems. For a quantitative measure of the degree of chaoticity of the many-body forces, the statistical distributions of the spacing between the nearest-neighboring levels were introduced in relation to the so called Random Matrix Theory (RMT). The fluctuation properties of quantum systems with underlying classical chaotic behavior and time reversal symmetry correspond with the predictions of the Gaussian orthogonal ensemble (GOE) of random matrix theory. On the contrary, integrable systems lead to level fluctuations that are well described by the Poisson distribution, i.e., levels behave as if they were uncorrelated [1-15]. The information on regular and chaotic nuclear motion available from experimental data is rather limited, because the analysis of energy levels requires the knowledge of sufficiently large pure sequences, i.e., consecutive level samples all with the same quantum numbers $(J,\pi)$ in a given nucleus. This means, one needs to combine different level schemes to prepare the sequences and perform a significant statistical study.

The rotational motion in rare earth nuclei which causes to band crossing, rotation-alignment and signature effects in the spectra and electromagnetic transitions and etc. [16-26], are considered as complicated phenomena which statistical method can be used to explain them. The effect of temperature on the spectral statistics of nuclear systems has been studied by in Ref.[16]. Also, the effect of spin on nuclear spectral statistics and also the shell correction as a function of angular momentum was calculated by extending the thermodynamical method for nonrotating nuclei to rotating nuclei [17].

In the present study, we have considered the statistical properties of even-even rare earth isotopes of *Ba*, *Ce*, *Nd*, *Sm*, *Gd*, *Dy*, *Er*, *Yb* and *Hf* nuclei which the spin-parity $J^{\pi}$ assignment of at least five consecutive levels are definite. We have focused on $0^+$, $2^+$ and $4^+$ levels of even-mass nuclei for their relative abundances. Sequences are prepared by using all the available empirical data [27-29] which are classified as their mass, quadrupole deformation parameter, decay modes, half-life amounts and shell model configuration for last protons and neutrons.

## 2. Method of analysis

Spectral fluctuations of nuclear levels have been analyzed by different statistics which are based on the comparison of statistical properties of nuclear spectra with the predictions of Random Matrix Theory (RMT) [1-2]. Nearest neighbor spacing distributions (NNSD), or *P*(s) functions, is the observable most commonly used to study the short-range fluctuation properties in nuclear spectra. NNSD statistics requires complete and pure level scheme which these condition are available for a limited number of nuclei. These requirements force us to combine level schemes of different nuclei to construct sequences. On the other hand, one must unfold the considered sequence which means each set of energy levels must be converted to a set of normalized spacing. To unfold our spectrum, we had to use some levels with same symmetry. This requirement is equivalent with the use of levels with same total quantum number (*J*) and same parity. For a given spectrum$\{E_i\}$, it is necessary to separate it into fluctuation part and smoothed average part, whose behavior is nonuniversal and cannot be described by RMT [1]. To do so, we count the number of the levels below *E* and write it as

$$N(E) = N_{av}(E) + N_{fluct}(E) \quad , \qquad (2.1)$$

Then we fix the $N_{ave}(E_i)$ semiclasically by taking a smooth polynomial function of degree 6 to fit the staircase function *N*(*E*). The unfolded spectrum is yield with the mapping

$$\{\tilde{E}_i\} = N(E_i) \quad , \quad (2.2)$$

This unfolded level sequence $\{\tilde{E}_i\}$, is dimensionless and has a constant average spacing of 1 but actual spacing exhibits frequently strong fluctuation. Nearest neighbor level spacing is defined as $s_i = (\tilde{E}_{i+1}) - (\tilde{E}_i)$. Distribution *P*(s) will be in such a way in which *P*(s)*d*s is the probability for the $s_i$ to lie within the infinitesimal interval [*s*,*s*+*ds*]. The NNS probability distribution function of nuclear systems which spectral spacing follows the Gaussian Orthogonal Ensemble (GOE) statistics is given by Wigner distribution [1]

$$P(s) = \frac{1}{2}\pi s e^{-\frac{\pi s^2}{4}} \quad , \quad (2.3)$$

this distribution exhibits the chaotic properties of spectra. On the other hand, the NNSD of systems with regular dynamics is generically represented by Poisson distribution

$$P(s) = e^{-s} \quad , \quad (2.4)$$

It is well known that real and complex systems such as nuclei are usually not fully ergodic and neither are they integrable. Different distribution functions have been proposed to compare the spectral statistics of considered systems with regular and chaotic limits quantitatively and also explore the interpolation between these limits [12-15]. Berry- Robnik distribution [13] is one of popular distribution

$$P(s,q) = [q + \frac{1}{2}\pi(1-q)s] \times \exp(-qs - \frac{1}{4}\pi(1-q)s^2) \quad , \quad (2.5)$$

is derived by assuming the energy level spectrum is a product of the superposition of independent subspectra which are contributed respectively from localized eigenfunctions into invariant (disjoint) phase space and interpolates between the Poisson and Wigner with $q = 1$ and 0, respectively. To consider the spectral statistics of sequences, one must compare it with Berry- Robnik distribution extract it’s parameter via estimation techniques. To avoid the disadvantages of estimation techniques such as Least square fitting (LSF) technique which has some unusual uncertainties for estimated values and also exhibit more approaches to chaotic dynamics, Maximum Likelihood (ML) technique has been used [10] which yields very exact results with low uncertainties in comparison with other estimation methods. The MLE estimation procedure has been described in detail in Refs.[10,15]. Here, we outline the basic ansatz and summarize the results. In order to estimate the parameter of distribution, Likelihood function is considered as product of all $P(s)$ functions,

$$L(q)=\prod_{i=1}^{n}P(s_i)=\prod_{i=1}^{n}[q+\frac{1}{2}\pi(1-q)s_i]\,e^{-qs_i-\frac{1}{4}\pi(1-q)s_i^2} \quad , \quad (2.6)$$

The desired estimator is obtained by maximizing the likelihood function, Eq.(1.6),

$$f{:}\sum\frac{1-\frac{\pi s_i}{2}}{q+\frac{\pi}{2}(1-q)s_i}-\sum(s_i-\frac{\pi s_i^2}{4}) \quad , \quad (2.7)$$

We can estimate "$q$ " by high accuracy via solving above equation by Newton-Raphson method

$$q_{new}=q_{old}-\frac{F(q_{old})}{F^{'}(q_{old})}$$

which is terminated to the following result,

$$q_{new}=q_{old}-\frac{\sum\frac{1-\frac{\pi s_i}{2}}{q_{old}+\frac{\pi s_i}{2}(1-q_{old})}+\sum s_i+\frac{\pi s_i^2}{4}}{\sum\frac{-(1-\frac{\pi s_i}{2})^2}{(q_{old}+\frac{1}{2}\pi(1-q_{old})s_i)^2}} \quad , \quad (2.8)$$

In ML-based technique, we have followed prescription explained in [10], namely maximum likelihood estimated parameters correspond to the converging values of iterations Eq. (2.8) which for the initial values we have chosen the values of parameters were obtained by LS method.

## 4. Results

In this paper, we consider the statistical properties of even-even rare earth nuclei. With regard to complete theoretical studies [30-41] and also the experimental evidences [31-33], we tend to classify nuclei in different categories. We considered all isotopes of *Ba*, *Ce*, *Nd*, *Sm*, *Gd*, *Dy*, *Er*, *Yb* and *Hf* nuclei which have at least five consecutive levels with definite spin-parity $J^{\pi}$ assignment as presented in Table 1. To test the effect due to mass, deformation and *etc.*, we have classified these nuclei in different sequences which are presented in Tables 2-5.

These sequences are unfolded and then analyzed via Berry-Robnik distribution and Maximum Likelihood estimation technique. Since, the exploration of the majority of short sequences yields an overestimation about the degree of chaotic dynamics which are measured by distribution parameters, i.e. $q$, therefore, we would not concentrate only on the implicit values of these quantities and examine a comparison between the amounts of this quantity in any table.

The ML-based predictions for Berry-Robnik distribution suggest a more regular dynamic for all sequences which are introduced in Table 2, as have presented in Figure 1. The apparent regularity for these deformed nuclei confirm the predictions of GOE limit which suggest more regular dynamics for deformed nuclei in comparison with the spherical nuclei, e.g. magic or semi magic. One can expect the spherical nuclei which have shell model spectra explore predominantly less regular dynamics in comparison with deformed ones. This result is known as AbulMagd-Weidenmuller chaocity effect [5] where suggest the suppression of chaotic dynamics due to the rotation of nuclei. Also, the more regularity for nuclei in $150< A< 180$ mass region in comparison with $100< A< 150$ region is similar to results are presented to Refs.[4,10].

In Table 3, we have classified the considered rare earth nuclei as their quadrupole deformation parameter and then determined their chaocity degrees via Berry-Robnik distribution and Maximum Likelihood technique. Our results explore a direct relation between deformation and regularity where the well deformed nuclei, $\beta_2 > 0.200$, explore the most regular dynamics. This result similar to results of Ref. [42] may be interpreted that, the degree of interaction between single particle motion which is chaotic and collective motion of whole nucleons which believed to be more regular is weaker in case of well deformed nuclei than other ones.

A comparison of the spacing distribution which are carried by the values of "$q$", ML-based predictions for Berry-Robnik distribution, for considered nuclei which are classified as their stability (or radioactivity), decay modes and half-life amounts are presented in Table 4. We have found stable nuclei explore more regular behavior in comparison with radioactive ones, which is similar to results of Ref.[43]. Also, the results show a dependence to decay modes and half-life amounts where radioactive nuclei which undergo through $\alpha$ decay suggest more chaotic dynamics in comparison to other radioactive nuclei. On the other hand, between different sequences of considered nuclei which classified as their half-life amounts, our results explore the more regular dynamics for a sequence with the shortest half-life. Although the lack of enough sample makes impossible to draw a conclusion, but it may indicate that something interesting happening in the structure of nuclei in relation to half-life. The majority of radioactive nuclei undergo through $\alpha$ decay have the longest half-life and also have spherical shapes but for a remarkable assumption we in need to consider more data.

In Table 5, we have used shell model configuration for level filling and classified our considered rare earth nuclei as their protons and neutrons configurations. In these categories, we look the dependence of spectral statistics to angular momentums, spins and shell effects which these procedures have carried in different references such as [31-40] with using theoretical predictions. Our results explore obvious dependence to the $J$ values of last proton and neutron levels. With increasing the $J$ values for neutron levels, $(1h_{9/2})^n \rightarrow (1h_{11/2})^n$ for a same proton level $(1g_{7/2})^p$ and $(2d_{5/2})^p$, or $(2f_{5/2})^n \rightarrow (2f_{7/2})^n$ for a same proton level $(1h_{11/2})^p$, the regularity of sequences are increased. On the contrary, we found an inverse relation between the increasing $J$ values for proton levels and chaocity. For a same neutron level $(1h_{9/2})^n$, when the $J$ values of proton level increased $(2d_{5/2})^p \rightarrow (1g_{7/2})^p$, or for a $(2f_{7/2})^n$ neutron level, when the $J$ values of proton level increased $(2d_{3/2})^p \rightarrow (1h_{11/2})^p$, our results obviously show that the chaocity degrees of sequences are increased. These results can be considered from the proton-proton or neutron-neutron interaction aspects. As have mentioned in Refs.[44-45], the relatively weak strength of the only neutron-neutron (or proton-proton) interaction is unable to destroy the regular single–particle mean–field motion completely. In some nuclei with both protons and neutrons in valence orbits, on the other hand, the stronger proton-neutron interaction would appear to be sufficient to destroy the regular mean–field motion. These mean, one may conclude that, when the $J$ values of neutrons are increased, the strength of interaction yield a more regular dynamics. This regularity also may be related to the strength of pairing force in comparison with Coulomb force but for a significant conclusion, we need to consider more general cases.

## 5. Summary and conclusion

We investigated the spectral statistics of even-even rare earth nuclei by using all the available experimental data. Berry- Robnik distribution and also MLE technique have been employed to consider the statistical situation of sequences. The difference in the chaoticity parameter of each sequence is statistically significant. Also, regular dynamics is dominant for well deformed nuclei in comparison with other ones. We have found an obvious relation between half-life amounts, stability (or radioactivity) and also the $J$ values of shell model configurations with the chaocity degrees of different sequences. The results show a deviation from GOE limit due to the increasing of the $J$ values in neutron levels which can realize as the effect of pairing force in nuclear structure. Also, these results may yield deep insight into the single-particle motion in the mean field formed by the deformed systems.

## Acknowledgement

This work is published as a part of research project supported by the University of Tabriz Research Affairs Office.

Table1. Shell model configuration for last protons and neutrons of considered nuclei. *N* denotes the number of levels, $\beta_2$ is the quadrupole deformation parameter and $T_{1/2}$ explores the half-life of considered nucleus. Also, the decay mode of radioactive nuclei is expressed in Table.

| Nuclei | $\beta_2$ | $T_{1/2}$ | decay mode | shell model configuration | levels | *N* | $E_{max}$ |
|---|---|---|---|---|---|---|---|
| $^{130}_{56}Ba$ | 0.215 | stable | $2\varepsilon$ | $(1g_{7/2})^p_6(1h_{11/2})^n_4$ | $2^+$ | 11 | 1943 |
| | | | | | $4^+$ | 5 | 2248 |
| $^{132}_{56}Ba$ | 0.185 | $>3\times10^{21}\ y$ | $\beta^-$ | $(1g_{7/2})^p_6(1h_{11/2})^n_6$ | $0^+$ | 6 | 2886 |
| | | | | | $2^+$ | 23 | 4028 |
| $^{134}_{56}Ba$ | 0.163 | stable | - | $(1g_{7/2})^p_6(1h_{11/2})^n_8$ | $0^+$ | 7 | 2784 |
| | | | | | $2^+$ | 29 | 3854 |
| | | | | | $4^+$ | 7 | 3079 |
| $^{136}_{56}Ba$ | 0.124 | stable | - | $(1g_{7/2})^p_6(1h_{11/2})^n_{10}$ | $0^+$ | 5 | 2784 |
| | | | | | $2^+$ | 19 | 3706 |
| $^{138}_{56}Ba$ | 0.093 | stable | - | $(1g_{7/2})^p_6(1h_{11/2})^n_{12}$ | $2^+$ | 29 | 3854 |
| $^{140}_{56}Ba$ | 0.116 | 12.75 $d$ | $\beta^-$ | $(1g_{7/2})^p_6(1h_{9/2})^n_2$ | $2^+$ | 13 | 3527 |
| $^{144}_{56}Ba$ | 0.196 | 11. 5 $s$ | $\beta^-$ | $(1g_{7/2})^p_6(1h_{9/2})^n_6$ | $2^+$ | 5 | 2357 |
| $^{146}_{56}Ba$ | 0.218 | 2.22 s | $\beta^-$ | $(1g_{7/2})^p_6(1h_{9/2})^n_8$ | $2^+$ | 6 | 3413 |
| $^{130}_{58}Ce$ | 0.257 | 22.9 $m$ | $\varepsilon$ | $(1g_{7/2})^p_8(1h_{11/2})^n_2$ | $2^+$ | 5 | 2116 |
| $^{132}_{58}Ce$ | 0.264 | 3.51 $h$ | $\varepsilon$ | $(1g_{7/2})^p_8(1h_{11/2})^n_4$ | $2^+$ | 13 | 2867 |
| $^{136}_{58}Ce$ | 0.264 | $>7\times10^{13}\ y$ | $2\varepsilon$ | $(1g_{7/2})^p_8(1h_{11/2})^n_8$ | $2^+$ | 6 | 2684 |
| $^{138}_{58}Ce$ | 0.126 | $>9\times10^{13}\ y$ | $2\varepsilon$ | $(1g_{7/2})^p_8(1h_{11/2})^n_{10}$ | $2^+$ | 10 | 3356 |
| | | | | | $4^+$ | 8 | 3082 |
| $^{140}_{58}Ce$ | 0.101 | stable | - | $(1g_{7/2})^p_8(1h_{11/2})^n_{12}$ | $0^+$ | 5 | 3226 |
| | | | | | $2^+$ | 7 | 3320 |
| | | | | | $4^+$ | 5 | 3521 |
| $^{142}_{58}Ce$ | 0.126 | $5\times10^{16}\ y$ | $2\beta^-$ | $(1g_{7/2})^p_8(1h_{9/2})^n_2$ | $2^+$ | 9 | 3697 |
| | | | | | $4^+$ | 7 | 2699 |
| $^{144}_{58}Ce$ | 0.162 | 284.9 $d$ | $\beta^-$ | $(1g_{7/2})^p_8(1h_{9/2})^n_4$ | $2^+$ | 11 | 2750 |
| $^{148}_{58}Ce$ | 0.253 | 56 $s$ | $\beta^-$ | $(1g_{7/2})^p_8(1h_{9/2})^n_8$ | $2^+$ | 5 | 1589 |
| $^{140}_{60}Nd$ | - | 3.37 $d$ | $\varepsilon$ | $(2d_{5/2})^p_2(1h_{11/2})^n_{10}$ | $0^+$ | 6 | 2911 |
| | | | | | $2^+$ | 13 | 3561 |
| | | | | | $4^+$ | 10 | 3493 |
| $^{142}_{60}Nd$ | 0.092 | stable | - | $(2d_{5/2})^p_2(1h_{11/2})^n_{12}$ | $0^+$ | 5 | 2983 |
| | | | | | $2^+$ | 6 | 3128 |
| $^{144}_{60}Nd$ | 0.134 | $2.29\times10^{15}\ y$ | $\alpha$ | $(2d_{5/2})^p_2(1h_{9/2})^n_2$ | $2^+$ | 7 | 2693 |
| | | | | | $4^+$ | 5 | 2401 |
| $^{146}_{60}Nd$ | 0.143 | stable | - | $(2d_{5/2})^p_2(1h_{9/2})^n_4$ | $0^+$ | 5 | 1697 |
| | | | | | $2^+$ | 6 | 1978 |
| | | | | | $4^+$ | 5 | 1989 |
| $^{148}_{60}Nd$ | 0.201 | stable | - | $(2d_{5/2})^p_2(1h_{9/2})^n_6$ | $0^+$ | 5 | 1600 |
| | | | | | $2^+$ | 11 | 2403 |
| | | | | | $4^+$ | 6 | 1683 |
| $^{142}_{62}Sm$ | - | $72.49\ m$ | $\varepsilon$ | $(2d_{5/2})^p_4(1h_{11/2})^n_{10}$ | $0^+$ | 6 | 2522 |
| | | | | | $2^+$ | 7 | 2374 |
| $^{146}_{62}Sm$ | - | $10.3\times10^7\ y$ | $\alpha$ | $(2d_{5/2})^p_4(1h_{9/2})^n_2$ | $2^+$ | 5 | 2544 |
| $^{148}_{62}Sm$ | 0.142 | $7\times10^{15}\ y$ | $\alpha$ | $(2d_{5/2})^p_4(1h_{9/2})^n_4$ | $2^+$ | 6 | 2146 |
| $^{150}_{62}Sm$ | 0.193 | stable | - | $(2d_{5/2})^p_4(1h_{9/2})^n_6$ | $2^+$ | 10 | 2055 |
| | | | | | $4^+$ | 10 | 2117 |
| $^{152}_{62}Sm$ | 0.308 | stable | - | $(2d_{5/2})^p_4(1h_{9/2})^n_8$ | $2^+$ | 6 | 2146 |
| | | | | | $4^+$ | 9 | 1901 |
| $^{154}_{62}Sm$ | 0.339 | stable | - | $(2d_{5/2})^p_4(1h_{9/2})^n_{10}$ | $2^+$ | 7 | 1879 |
| $^{146}_{64}Gd$ | - | $48.27\ d$ | $\varepsilon$ | $(2d_{5/2})^p_6(1h_{11/2})^n_{10}$ | $0^+$ | 5 | 3639 |
| | | | | | $2^+$ | 7 | 3552 |
| $^{148}_{64}Gd$ | 0.077 | $70.9\ y$ | $\alpha$ | $(2d_{5/2})^p_6(1h_{9/2})^n_2$ | $2^+$ | 7 | 2700 |
| $^{150}_{64}Gd$ | 0.077 | $1.79\times10^6\ y$ | $\alpha$ | $(2d_{5/2})^p_6(1h_{9/2})^n_4$ | $2^+$ | 5 | 1988 |
| $^{152}_{64}Gd$ | 0.212 | $1.08\times10^{14}\ y$ | $\alpha$ | $(2d_{5/2})^p_6(1h_{9/2})^n_6$ | $2^+$ | 7 | 1471 |
| $^{154}_{64}Gd$ | 0.310 | stable | - | $(2d_{5/2})^p_6(1h_{9/2})^n_8$ | $0^+$ | 9 | 1899 |
| | | | | | $2^+$ | 11 | 2024 |
| | | | | | $4^+$ | 6 | 1789 |
| $^{156}_{64}Gd$ | 0.340 | stable | - | $(2d_{5/2})^p_6(1h_{9/2})^n_{10}$ | $0^+$ | 5 | 1851 |
| | | | | | $2^+$ | 10 | 2217 |
| | | | | | $4^+$ | 6 | 1893 |
| $^{158}_{64}Gd$ | 0.349 | stable | - | $(2d_{5/2})^p_6(2f_{7/2})^n_2$ | $0^+$ | 6 | 1958 |
| | | | | | $2^+$ | 12 | 2260 |
| | | | | | $4^+$ | 8 | 2030 |
| $^{160}_{64}Gd$ | 0.348 | $>3.1\times10^{19}\ y$ | $2\beta^-$ | $(2d_{5/2})^p_6(2f_{7/2})^n_4$ | $2^+$ | 11 | 1599 |
| | | | | | $4^+$ | 8 | 2121 |
| $^{154}_{66}Dy$ | 0.235 | $3\times10^6\ y$ | $\alpha$ | $(2d_{3/2})^p_2(1h_{9/2})^n_6$ | $2^+$ | 5 | 1507 |
| $^{156}_{66}Dy$ | 0.294 | stable | - | $(2d_{3/2})^p_2(1h_{9/2})^n_8$ | $2^+$ | 9 | 2090 |
| | | | | | $4^+$ | 5 | 1677 |
| $^{158}_{66}Dy$ | 0.326 | stable | - | $(2d_{3/2})^p_2(1h_{9/2})^n_{10}$ | $0^+$ | 5 | 1743 |
| | | | | | $2^+$ | 10 | 1975 |
| | | | | | $4^+$ | 9 | 2056 |
| $^{160}_{66}Dy$ | 0.334 | stable | - | $(2d_{3/2})^p_2(2f_{7/2})^n_2$ | $2^+$ | 5 | 1557 |
| | | | | | $4^+$ | 6 | 1652 |
| $^{162}_{66}Dy$ | 0.341 | stable | - | $(2d_{3/2})^p_2(2f_{7/2})^n_4$ | $2^+$ | 6 | 1895 |
| | | | | | $4^+$ | 5 | 1887 |
| $^{164}_{66}Dy$ | 0.347 | stable | - | $(2d_{3/2})^p_2(2f_{7/2})^n_6$ | $2^+$ | 5 | 1738 |
| $^{156}_{68}Er$ | 0.189 | $19.5\ m$ | $\varepsilon$ | $(2d_{3/2})^p_4(1h_{9/2})^n_6$ | $2^+$ | 5 | 1910 |
| $^{158}_{68}Er$ | 0.264 | 2.29 $h$ | $\varepsilon$ | $(2d_{3/2})^p_4(1h_{9/2})^n_8$ | $2^+$ | 12 | 1698 |
| | | | | | $4^+$ | 5 | 1427 |
| $^{160}_{68}Er$ | 0.300 | 28.58 $h$ | $\varepsilon$ | $(2d_{3/2})^p_4(1h_{9/2})^n_{10}$ | $2^+$ | 5 | 1390 |
| $^{166}_{68}Er$ | 0.344 | stable | - | $(2d_{3/2})^p_4(2f_{7/2})^n_6$ | $2^+$ | 5 | 1894 |
| $^{168}_{68}Er$ | 0.340 | stable | - | $(2d_{3/2})^p_4(2f_{7/2})^n_8$ | $2^+$ | 5 | 1812 |
| | | | | | $4^+$ | 5 | 1736 |
| $^{170}_{68}Er$ | 0.336 | stable | - | $(2d_{3/2})^p_4(2f_{5/2})^n_2$ | $2^+$ | 6 | 1416 |
| | | | | | $4^+$ | 8 | 1573 |
| $^{160}_{70}Yb$ | 0.219 | 4.8 $m$ | $\varepsilon$ | $(3s_{1/2})^p_2(1h_{9/2})^n_8$ | $2^+$ | 8 | 1811 |
| $^{162}_{70}Yb$ | 0.263 | 18.87 $m$ | $\varepsilon$ | $(3s_{1/2})^p_2(1h_{9/2})^n_{10}$ | $2^+$ | 8 | 1398 |
| $^{164}_{70}Yb$ | 0.296 | 75.8 $m$ | $\varepsilon$ | $(3s_{1/2})^p_2(1f_{7/2})^n_2$ | $2^+$ | 7 | 1513 |
| | | | | | $4^+$ | 6 | 1612 |
| $^{168}_{70}Yb$ | 0.327 | stable | - | $(3s_{1/2})^p_2(2f_{7/2})^n_6$ | $2^+$ | 5 | 1604 |
| $^{170}_{70}Yb$ | 0.321 | stable | - | $(3s_{1/2})^p_2(2f_{7/2})^n_8$ | $0^+$ | 5 | 1566 |
| | | | | | $2^+$ | 6 | 1658 |
| $^{172}_{70}Yb$ | 0.330 | stable | - | $(3s_{1/2})^p_2(2f_{5/2})^n_2$ | $2^+$ | 5 | 1608 |
| | | | | | $4^+$ | 8 | 1803 |
| $^{166}_{72}Hf$ | 0.249 | 6.77 $m$ | $\varepsilon$ | $(1h_{11/2})^p_2(2f_{7/2})^n_2$ | $2^+$ | 5 | 1603 |
| $^{168}_{72}Hf$ | 0.273 | 25.95 $m$ | $\varepsilon$ | $(1h_{11/2})^p_2(2f_{7/2})^n_4$ | $2^+$ | 6 | 1800 |
| $^{172}_{72}Hf$ | 0.296 | 1.87 $y$ | $\varepsilon$ | $(1h_{11/2})^p_2(2f_{7/2})^n_8$ | $2^+$ | 7 | 1575 |
| | | | | | $4^+$ | 10 | 1601 |
| $^{174}_{72}Hf$ | 0.272 | $3\times10^{15}$ $y$ | $\alpha$ | $(1h_{11/2})^p_2(2f_{5/2})^n_2$ | $2^+$ | 14 | 2530 |
| $^{176}_{72}Hf$ | 0.298 | stable | - | $(1h_{11/2})^p_2(2f_{5/2})^n_4$ | $2^+$ | 5 | 1692 |
| $^{178}_{72}Hf$ | 0.279 | stable | - | $(1h_{11/2})^p_2(2f_{5/2})^n_6$ | $2^+$ | 9 | 1891 |
| | | | | | $4^+$ | 8 | 1870 |

Table 2. The chaocity parameters, "*q*" Berry-Robnik distribution parameter, are determined for different sequences which are classified as their masses and then analyzed via MLE technique. *N* is the number of spacing.

| Sequence | *all levels* | | *only* $0^+$ *levels* | | *only* $2^+$ *levels* | | *only* $4^+$ *levels* | |
|---|---|---|---|---|---|---|---|---|
| | *N* | *q* | *N* | *q* | *N* | *q* | *N* | *q* |
| 100< A< 150 | 361 | 0.68±0.09 | 49 | 0.53±0.12 | 260 | 0.71±0.08 | 52 | 0.51±0.16 |
| 150< A< 180 | 322 | 0.76±0.11 | 28 | 0.59±0.06 | 195 | 0.80±0.10 | 99 | 0.55±0.14 |

Table 3. The chaocity parameters, "$q$" Berry-Robnik distribution parameter, are determined for different sequences which are classified as their quadrupole deformation parameter and then analyzed via MLE technique. $N$ is the number of spacing which includes all $0^+$, $2^+$ and $4^+$ levels.

| Sequence | $N$ | $q$ |
|---|---|---|
| $0.077 < \beta_2 < 0.126$ | 115 | 0.67±0.08 |
| $0.134 < \beta_2 < 0.196$ | 131 | 0.72±0.14 |
| $0.201 < \beta_2 < 0.296$ | 173 | 0.78±0.06 |
| $0.308 < \beta_2 < 0.349$ | 211 | 0.83±0.10 |

Table 4. The chaocity parameters, "$q$" Berry-Robnik distribution parameter, are determined for different sequences which are classified as their stability and radioactivity modes and also their $T_{1/2}$ values. $N$ is the number of spacing which includes all $0^+$, $2^+$ and $4^+$ levels.

| Sequence | $N$ | $q$ |
|---|---|---|
| Stable nuclei (27 nuclei) | 369 | 0.72±0.13 |
| All radioactive nuclei (32 nuclei) | 261 | 0.38±0.08 |
| Radioactive nuclei undergo through $\varepsilon$ decay | 127 | 0.44±0.09 |
| Radioactive nuclei undergo through $\beta^-$ decay | 93 | 0.58±0.10 |
| Radioactive nuclei undergo through $\alpha$ decay | 53 | 0.31±0.05 |
| Radioactive nuclei with $T_{1/2}\sim m$ | 49 | 0.37±0.10 |
| Radioactive nuclei with $T_{1/2}\sim d$ | 58 | 0.33±0.11 |
| Radioactive nuclei with $T_{1/2}\sim h$ | 31 | 0.29±0.05 |
| Radioactive nuclei with $T_{1/2}\sim y$ ($1.87\, y < T_{1/2} < 3\times10^{21}\, y$) | 146 | 0.25±0.07 |

Table 5. The chaocity parameters, "$q$" Berry-Robnik distribution parameter, are determined for different sequences which are classified as their shell model configuration for last protons and neutrons. $N$ is the number of spacing which includes all $0^+$, $2^+$ and $4^+$ levels.

| Sequence | $N$ | $q$ |
|---|---|---|
| Nuclei with $(1g_{7/2})^p_{6-8}(1h_{11/2})^n_{2-12}$ configuration | 182 | 0.63±0.08 |
| Nuclei with $(1g_{7/2})^p_{6-8}(1h_{9/2})^n_{2-8}$ configuration | 49 | 0.54±0.09 |
| Nuclei with $(2d_{5/2})^p_{2-6}(1h_{11/2})^n_{10-12}$ configuration | 56 | 0.71±0.11 |
| Nuclei with $(2d_{5/2})^p_{2-6}(1h_{9/2})^n_{2-10}$ configuration | 80 | 0.66±0.14 |
| Nuclei with $(2d_{5/2})^p_{6}(2f_{7/2})^n_{2-4}$ configuration | 39 | 0.53±0.10 |
| Nuclei with $(2d_{3/2})^p_{2-4}(1h_{9/2})^n_{6-10}$ configuration | 60 | 0.73±0.04 |
| Nuclei with $(2d_{3/2})^p_{2-4}(2f_{7/2})^n_{2-8}$ configuration | 34 | 0.52±0.06 |
| Nuclei with $(3s_{1/2})^p_{2}(2f_{7/2})^n_{2-8}$ configuration | 24 | 0.64±0.05 |
| Nuclei with $(1h_{11/2})^p_{2}(2f_{7/2})^n_{2-8}$ configuration | 24 | 0.25±0.11 |
| Nuclei with $(1h_{11/2})^p_{2}(2f_{5/2})^n_{2-6}$ configuration | 34 | 0.21±0.10 |

# Figure caption

Figure1. NNSD histograms are presented for two sequences of Table 1 which contains all levels of these two mass regions. Solid, dashed and dotted line represent the Poisson, GOE and Berry-Robnik distribution curves, respectively.

Fig1.

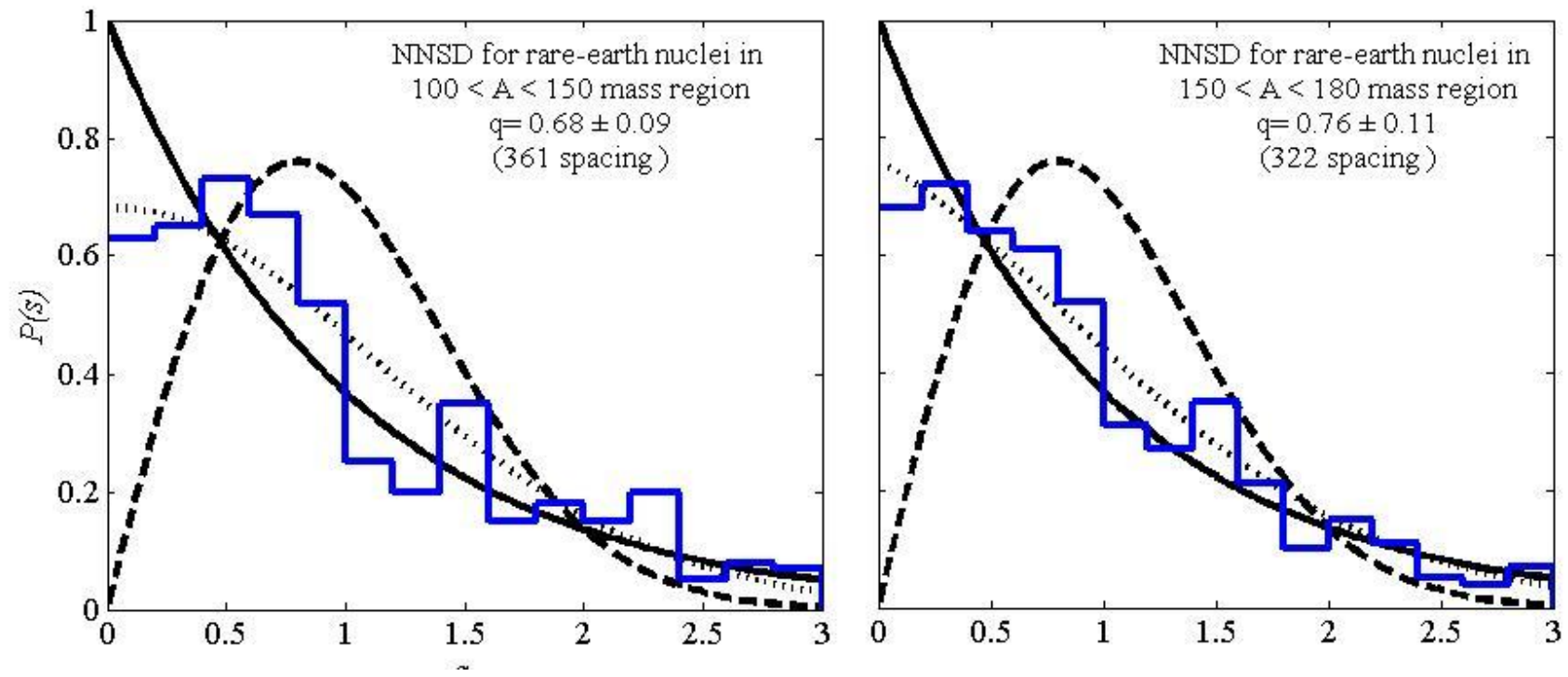